\documentclass[12pt]{article}

\usepackage{axodraw}
\usepackage{epsfig}

\makeatletter

\def\paragraph{\@startsection{paragraph}{4}{\z@}{+2.00ex plus
 +1ex minus +.2ex}{1.5ex plus .2ex}{\it\normalsize}}

\def\section{\@startsection {section}{1}{\z@}{+3.0ex plus +1ex minus
  +.2ex}{2.3ex plus .2ex}{\normalsize\bf\boldmath}}
\def\subsection{\@startsection{subsection}{2}{\z@}{+2.5ex plus +1ex
minus +.2ex}{1.5ex plus .2ex}{\normalsize\bf\boldmath}}
\def\subsubsection{\@startsection{subsubsection}{3}{\z@}{+3.25ex plus
 +1ex minus +.2ex}{1.5ex plus .2ex}{\normalsize\it}}

\expandafter\ifx\csname mathrm\endcsname\relax\def\mathrm#1{{\rm #1}}\fi

\newcounter{saveeqn}

\@addtoreset{equation}{section}

\newcount\@tempcntc
\def\@citex[#1]#2{\if@filesw\immediate\write\@auxout{\string\citation{#2}}\fi
  \@tempcnta\z@\@tempcntb\m@ne\def\@citea{}\@cite{\@for\@citeb:=#2\do
    {\@ifundefined
       {b@\@citeb}{\@citeo\@tempcntb\m@ne\@citea
        \def\@citea{,\penalty\@m\ }{\bf ?}\@warning
       {Citation `\@citeb' on page \thepage \space undefined}}%
    {\setbox\z@\hbox{\global\@tempcntc0\csname
b@\@citeb\endcsname\relax}%
     \ifnum\@tempcntc=\z@ \@citeo\@tempcntb\m@ne
       \@citea\def\@citea{,\penalty\@m}
       \hbox{\csname b@\@citeb\endcsname}%
     \else
      \advance\@tempcntb\@ne
      \ifnum\@tempcntb=\@tempcntc
      \else\advance\@tempcntb\m@ne\@citeo
      \@tempcnta\@tempcntc\@tempcntb\@tempcntc\fi\fi}}\@citeo}{#1}}

\def\@citeo{\ifnum\@tempcnta>\@tempcntb\else\@citea
  \def\@citea{,\penalty\@m}%
  \ifnum\@tempcnta=\@tempcntb\the\@tempcnta\else
   {\advance\@tempcnta\@ne\ifnum\@tempcnta=\@tempcntb \else
\def\@citea{--}\fi
    \advance\@tempcnta\m@ne\the\@tempcnta\@citea\the\@tempcntb}\fi\fi}

\def\nl{\nonumber\\}

\newcommand{\gsim}
{\mathrel{\raisebox{-.3em}{$\stackrel{\displaystyle >}{\sim}$}}}
\def\asymp#1%
{\mathrel{\raisebox{-.4em}{$\widetilde{\scriptstyle #1}$}}}

\def\Nequal#1%
{\mathrel{\raisebox{-.5em}{$\stackrel{=}{\scriptstyle\rm#1}$}}}
\newcommand{\dsl}[1]{\not \hspace{-0.7mm}#1}
\def\dsl{\mathpalette\make@slash}
\def\make@slash#1#2{\setbox\z@\hbox{$#1#2$}%
  \hbox to 0pt{\hss$#1/$\hss\kern-\wd0}\box0}

\def\beq{\begin{equation}}
\def\eeq{\end{equation}}
\def\beqar{\begin{eqnarray}}
\def\eeqar{\end{eqnarray}}
\def\barr#1{\begin{array}{#1}}
\def\earr{\end{array}}
\def\bfi{\begin{figure}}
\def\efi{\end{figure}}
\def\btab{\begin{table}}
\def\etab{\end{table}}
\def\bce{\begin{center}}
\def\ece{\end{center}}

\def\text{\textstyle}

\def\al{\alpha}

\def\ga{\gamma}
\def\de{\delta}

\def\si{\sigma}

\def\refeq#1{\mbox{(\ref{#1})}}

\def\reffi#1{\mbox{Figure~\ref{#1}}}
\def\reffis#1{\mbox{Figures~\ref{#1}}}
\def\refta#1{\mbox{Table~\ref{#1}}}

\def\citere#1{\mbox{Ref.~\cite{#1}}}
\def\citeres#1{\mbox{Refs.~\cite{#1}}}

\newcommand{\GeV}{\unskip\,\mathrm{GeV}}
\newcommand{\MeV}{\unskip\,\mathrm{MeV}}
\newcommand{\keV}{\unskip\,\mathrm{keV}}

\newcommand{\fb}{\unskip\,\mathrm{fb}}

\newcommand{\rd}{{\mathrm{d}}}

\newcommand{\Oa}{\mathswitch{{\cal{O}}(\alpha)}}

\def\mathswitchr#1{\relax\ifmmode{\mathrm{#1}}\else$\mathrm{#1}$\fi}

\newcommand{\PW}{\mathswitchr W}
\newcommand{\Pw}{\mathswitchr w}
\newcommand{\PZ}{\mathswitchr Z}

\newcommand{\PH}{\mathswitchr H}
\newcommand{\Pe}{\mathswitchr e}

\newcommand{\Pd}{\mathswitchr d}
\newcommand{\Pdbar}{\bar{\mathswitchr d}}
\newcommand{\Pu}{\mathswitchr u}
\newcommand{\Pubar}{\bar{\mathswitchr u}}
\newcommand{\Ps}{\mathswitchr s}

\newcommand{\Pcbar}{\bar{\mathswitchr c}}

\newcommand{\Pt}{\mathswitchr t}

\newcommand{\Pep}{\mathswitchr {e^+}}
\newcommand{\Pem}{\mathswitchr {e^-}}
\newcommand{\PWp}{\mathswitchr {W^+}}
\newcommand{\PWm}{\mathswitchr {W^-}}

\def\mathswitch#1{\relax\ifmmode#1\else$#1$\fi}

\newcommand{\MW}{\mathswitch {M_\PW}}
\newcommand{\MWfit}{\mathswitch {M_{\PW,\mathrm{fit}}}}

\newcommand{\MZ}{\mathswitch {M_\PZ}}
\newcommand{\MH}{\mathswitch {M_\PH}}
\newcommand{\Me}{\mathswitch {m_\Pe}}

\newcommand{\Mt}{\mathswitch {m_\Pt}}
\newcommand{\GW}{\Gamma_{\PW}}
\newcommand{\GWfit}{\mathswitch {\Gamma_{\PW,\mathrm{fit}}}}
\newcommand{\GZ}{\Gamma_{\PZ}}

\newcommand{\sw}{\mathswitch {s_\Pw}}
\newcommand{\cw}{\mathswitch {c_\Pw}}

\newcommand{\GF}{\mathswitch {G_\mu}}

\def\solid{\raise.9mm\hbox{\protect\rule{1.1cm}{.2mm}}}
\def\dash{\raise.9mm\hbox{\protect\rule{2mm}{.2mm}}\hspace*{1mm}}

\def\ie{i.e.\ }

\newcommand{\born}{{\mathrm{Born}}}
\newcommand{\corr}{{\mathrm{corr}}}

\newcommand{\eeWWffff}{\Pep\Pem\to\PW\PW\to 4f}
\newcommand{\eeWWffffg}{\eeWWffff\gamma}
\newcommand{\eeffff}{\Pep\Pem\to 4f}
\newcommand{\eeffffg}{\eeffff\ga}
\renewcommand{\O}{{\cal O}}

\def\draftdate{\relax}
\def\mda{\relax}
\def\mua{\relax}
\def\mla{\relax}
\def\draft{
\def\thtystars{******************************}
\def\sixtystars{\thtystars\thtystars}
\typeout{}
\typeout{\sixtystars**}
\typeout{* Draft mode!
         For final version remove \protect\draft\space in source file *}
\typeout{\sixtystars**}
\typeout{}
\def\draftdate{\today}
\def\mua{\marginpar[\boldmath\hfil$\uparrow$]%
                   {\boldmath$\uparrow$\hfil}%
                    \typeout{marginpar: $\uparrow$}\ignorespaces}
\def\mda{\marginpar[\boldmath\hfil$\downarrow$]%
                   {\boldmath$\downarrow$\hfil}%
                    \typeout{marginpar: $\downarrow$}\ignorespaces}
\def\mla{\marginpar[\boldmath\hfil$\rightarrow$]%
                   {\boldmath$\leftarrow $\hfil}%
                    \typeout{marginpar: $\leftrightarrow$}\ignorespaces}
\def\Mua{\marginpar[\boldmath\hfil$\Uparrow$]%
                   {\boldmath$\Uparrow$\hfil}%
                    \typeout{marginpar: $\uparrow$}\ignorespaces}
\def\Mda{\marginpar[\boldmath\hfil$\Downarrow$]%
                   {\boldmath$\Downarrow$\hfil}%
                    \typeout{marginpar: $\downarrow$}\ignorespaces}
\def\Mla{\marginpar[\boldmath\hfil$\Rightarrow$]%
                   {\boldmath$\Leftarrow $\hfil}%
                    \typeout{marginpar: $\leftrightarrow$}\ignorespaces}
\overfullrule 5pt
\oddsidemargin -15mm
\marginparwidth 29mm
}

\def\stars{\strut\leaders\hbox{*}\hfill\strut}
\def\starline{\hfil\strut\hfil\hbox to \textwidth {\stars}\hfil}

\begin{document}
\thispagestyle{empty}
\def\thefootnote{\fnsymbol{footnote}}
\setcounter{footnote}{1}
\null
\draftdate\hfill BI-TP 99/47 \\
\strut\hfill ER/40685/941\\
\strut\hfill LU-ITP 1999/023\\
\strut\hfill PSI-PR-99-34\\
\strut\hfill UR-1594\\
\strut\hfill hep-ph/9912447
\vfill
\begin{center}
{\Large \bf\boldmath
W-pair production at future $\Pep\Pem$ colliders:
\\[.5em]
precise predictions from {\sc RacoonWW}
\par} \vskip 2.5em
\vspace{1cm}

{\large
{\sc A.\ Denner$^1$, S.\ Dittmaier$^2$, M. Roth$^{3}$ and 
D.\ Wackeroth$^4$} } \\[1cm]

$^1$ {\it Paul Scherrer Institut\\
CH--5232 Villigen PSI, Switzerland} \\[0.5cm]

$^2$ {\it Theoretische Physik, Universit\"at Bielefeld \\
D--33615 Bielefeld, Germany}
\\[0.5cm]

$^3$ {\it Institut f\"ur Theoretische Physik, Universit\"at Leipzig\\
D--04109 Leipzig, Germany}
\\[0.5cm]

$^4$ {\it Department of Physics and Astronomy, University of Rochester\\
Rochester, NY 14627-0171, USA}
\par \vskip 1em
\end{center}\par
\vskip 2cm {\bf Abstract:} 
\par 
We present numerical results for total cross sections and various 
distributions for $\eeWWffff(+\gamma)$ at a future $500\GeV$ linear
collider, obtained from the Monte Carlo generator {\sc RacoonWW}.
This generator is the first one that includes $\O(\al)$ electroweak 
radiative corrections in the double-pole approximation completely. 
Owing to their large size the corrections are of great phenomenological
importance.
\par
\vskip 1cm
\noindent
December 1999
\null
\setcounter{page}{0}
\clearpage
\def\thefootnote{\arabic{footnote}}
\setcounter{footnote}{0}

One of the most important class of processes to be investigated at
future 
$\Pep\Pem$ linear colliders are the four-fermion production processes,
which involve in particular W-pair production, $\eeWWffff$.  In order
to match the experimental accuracy for this process of 1\% or better,
theoretical predictions for the corresponding cross sections of
$\eeffff$ with a precision at the level of some 0.1\% are needed. This
requires to include the complete set of lowest-order diagrams for
$\eeffff$ and the $\Oa$ corrections to the W-pair production channels
$\eeWWffff$. In addition, the leading higher-order corrections should
be taken into account.

In all regions of phase space where \PW-pair production dominates the
cross section for $\eeffff$, an expansion of the matrix element about
the poles of the resonant \PW~propagators provides a reasonable
approximation for the radiative corrections.  Neglecting corrections
to the non-doubly-resonant contributions leads to uncertainties of the
order $\al/\pi\times\GW/\MW\times\log(\ldots)\sim 0.1\%$ with respect
to the leading lowest-order contributions, where the logarithm
indicates possible logarithmic enhancements.  
This so-called double-pole approximation (DPA)
provides a gauge-invariant answer and allows us to
use the existing results for on-shell W-pair production
\cite{rcwprod1,rcwprod2} and \PW-boson decay
\cite{rcwdecay1,rcwdecay2}, as far as the virtual corrections are
concerned.

The DPA for $\Oa$ corrections to pair production of unstable particles
has already been proposed in \citere{Ae94}. Recently different variants
of the DPA have been used in the literature 
\cite{Be98,ja97,ja99,ku99,racoonww1}.
A Monte Carlo
generator including the corrections to the \PW-pair production
subprocess and the leading-logarithmic corrections to the \PW-boson
decays has been constructed in 
\citeres{ja97,ja99}, but non-factorizable
corrections and W-spin correlations have been neglected there. With
this generator also results for  the centre-of-mass (CM) energy 
$\sqrt{s}=500\GeV$ have been
produced. A different, but also approximate, implementation 
\cite{ku99} of the DPA has been compared at
energies $\sqrt{s}\gsim500\GeV$ with a high-energy approximation 
\cite{be93} for $\eeWWffff$. The first complete calculation of the $\Oa$
corrections for off-shell \PW-pair production, including a numerical
study of leptonic final states for LEP2 energies, was presented in
\citere{Be98} using a semi-analytical approach, which is, however,
only applicable to ideal theoretical situations.

Recently the first Monte-Carlo generator that incorporates the
complete $\Oa$ corrections for off-shell \PW-pair production in DPA
has been constructed. This generator, called {\sc RacoonWW}
\cite{racoonww1,racoonww2}, includes the complete lowest-order matrix
elements for $\eeffff$ for any four-fermion final state. For the
virtual corrections a DPA is used without any additional
approximations. In particular, the exact four-fermion phase space is
used throughout. The virtual corrections consist of factorizable and
non-factorizable contributions. The former are the ones that are
associated to either \PW-pair production or \PW-boson decay; the
results of \citeres{rcwprod1,rcwdecay1} are used in this part. The
latter comprise all corrections in which the subprocesses production
and decays do not proceed independently. Up to some simple
supplements, the virtual non-factorizable corrections can be read off
from the literature \cite{nfc1,nfc2}; we made use of the results of
\citere{nfc2} and included the exact off-shell Coulomb singularity
\cite{Coul}. The real bremsstrahlung corrections are based on the full
matrix-element calculation for $\eeffffg$ described in \citere{ee4fa}.
More precisely, the minimal gauge-invariant subset including all
doubly-resonant contributions of the processes $\eeWWffffg$, \ie the
photon radiation from the CC11 subset, are included.  By using the
exact matrix elements for the real radiation, we avoid problems in
defining a DPA for semi-soft photons ($E_\gamma\sim\GW$) and,
moreover, can include the leading logarithmic corrections to
non-doubly-resonant diagrams (background diagrams) exactly.  The real
corrections and the virtual corrections are matched in such a way that
all infrared singularities cancel exactly.  The initial-state
collinear singularities are regularized by retaining a finite electron
mass and factorized into lowest-order matrix element and splitting
functions. The collinear singularities connected to final-state
radiation are treated inclusively, \ie photons within collinear cones
around the final-state fermions are integrated over, so that no
logarithmic final-state fermion mass dependence survives.  All
contributions have been implemented in two programs, one of which uses
the subtraction method described in \citere{subtract}, the other one
uses phase-space slicing. All parts of the calculations have been
performed in two independent ways.  A detailed description of the
calculation and the Monte Carlo generator {\sc RacoonWW} will be
published elsewhere \cite{racoonww2}. Results for the LEP2-energy
region have already been given in \citere{racoonww1}. Here we
concentrate on results for $\sqrt{s}=500\GeV$,
which is relevant for a future linear collider. \unskip  
\bigskip

For the numerical evaluation we used the same input as in
\citere{racoonww1}, i.e.\ the fixed-width scheme and the
following parameters:
\beq
\begin{array}[b]{rlrl}
\GF =& 1.16637\times 10^{-5} \GeV^{-2}, \qquad&\alpha=&1/137.0359895, \\
\MW =& 80.35\GeV,& \GW =& 2.08699\ldots\GeV, \\
\MZ =& 91.1867\GeV,& \GZ =& 2.49471\GeV, \\
\Mt =& 174.17 \GeV,&\MH=& 150\GeV, \\
\Me =& 510.99907 \keV.
\end{array}
\label{eq:input}
\eeq
The reliability of the
fixed-width scheme at high energies has been demonstrated
in \citere{bhf}.
The weak mixing angle is fixed by $\cw=\MW/\MZ$, $\sw^2=1-\cw^2$.
The parameters in \refeq{eq:input} 
are over-complete but self-consistent. Instead
of $\alpha$ we use $\GF$ to parameterize the lowest-order matrix
element, \ie we use the effective coupling
\beq
\alpha_{\GF} = \frac{\sqrt{2}\GF\MW^2\sw^2}{\pi}
\eeq
in the lowest-order matrix element. This parameterization has the
advantage that all higher-order contributions associated with the
running of the electromagnetic coupling and the leading universal
two-loop $\Mt$-dependent corrections to the dominant contributions are
correctly taken into account.  In the relative $\Oa$ corrections, on
the other hand, we use $\al$, since in the real corrections the scale
of the real photon is zero.  The W-boson width given above is
calculated including the electroweak and QCD one-loop corrections with
$\alpha_{\mathrm s}=0.119$.  We do not include QCD corrections to the
process $\eeWWffff$, and initial-state radiation is only taken into
account in $\Oa$.

In \refta{table500a}  we present
numbers for total cross sections without any cuts based on 20
million events. In particular, we give the CC03 cross sections, \ie
the cross sections resulting from the signal diagrams only (defined in
the 't~Hooft--Feynman gauge). We also give numbers resulting from the
complete set of diagrams for those final states where this is possible
without cuts, \ie the CC11 class of processes.  In the considered
cases, the effects of the background diagrams are below $0.6\%$.  Note
that we treat the external fermions as massless. Therefore, the cross
sections for processes not in CC11 class become singular if no cuts
are imposed for final-state electrons collinear to the beams and for
virtual photons splitting into $f\bar f$ pairs with small invariant
masses.
\begin{table}
$$
\begin{array}{l@{\qquad}c@{\qquad}c@{\qquad}c@{\qquad}c}
\hline
\mbox{final state} &
\mbox{CC03 Born} &  \mbox{full Born} & 
\mbox{CC03 corrected} & \mbox{full corrected} \nl
\hline
\nu_\mu \mu^+ \tau^-\bar\nu_\tau &
 87.12(4)  &  86.66(3) & 90.27(4)  &  89.81(4) \nl  
&87.11(4)  &  86.65(3) & 90.19(5)  &  89.73(5) \nl  
\nu_\mu\mu^+\Pd\Pubar   & 
 261.4(1)  &  260.0(1)  & 270.6(1) &  269.3(1) \nl  
&261.3(1)  &  260.0(1)  & 270.3(2) &  269.0(2) \nl  
\Pu\Pdbar\Ps\Pcbar & 
 784.1(3)  &  780.3(3) & 811.3(4)  &  807.5(4) \nl  
&784.0(3)  &  780.4(3) & 810.9(4)  &  807.2(4) \nl  
\mbox{total} &
  7057(2) & & 7305(2) \nl 
& 7056(2) & & 7299(3) \nl 
\hline
\end{array}
$$
\caption{Total cross sections in fb for $\eeWWffff$ without cuts 
for various final states at $\protect\sqrt{s}=500\GeV$}
\label{table500a}
\end{table}
The shown corrections are typically $3.5\%$ at $\sqrt{s}=500\GeV$. The
numbers in parentheses are estimates of the Monte Carlo integration
errors.  The numbers for the total \PW-pair (CC03) production cross
section in the
last two rows of the tables directly result from the other rows by
multiplying these with the number of equivalent channels and adding
them up.  The numbers in the upper lines are from the
subtraction-method branch of {\sc RacoonWW}, the number in the lower
lines from the phase-space slicing branch. The agreement between these
numbers serves as a stringent consistency check of our generator.

Note that the total CC03 cross section of \refta{table500a} is different 
from the full cross section for four-fermion production, as 
non-doubly-resonant (background) diagrams significantly contribute to 
the total cross section for high energies. The most important class of 
background diagrams are the ones for single-W production, which is 
dominant if electrons or positrons are scattered in the very forward 
directions (see e.g.\ \citere{bo99} and references therein). 
The DPA approach is, of course, restricted to situations where the 
doubly-resonant diagrams dominate the cross section. Therefore, before
confronting the ${\cal O}(\alpha)$-corrected prediction of {\sc RacoonWW} 
with empirical data one has either to specify cuts that suppress 
background diagrams or to extract the W-pair signal from the full 
four-fermion signature. 

Next, we study observables that involve phase-space and 
photon-recombination cuts.
Because of our treatment of mass singularities 
(see \citere{racoonww2} for details) it is
necessary to combine photons that are collinear to the incoming or
outgoing fermions appropriately with these fermions in order to obtain
well-defined finite distributions.
To this end we introduce the following recombination and cut procedure
which proceeds in three steps:
\newcommand{\recomb}{{\mathrm{rec}}}%
\begin{enumerate}
\item All photons within a cone of 5 degrees around the beams are
  treated as invisible, \ie their momenta are disregarded when
  calculating angles, energies,
  and invariant masses.
\item Next, the invariant masses of the photon with 
each of the charged final-state fermions are calculated. If the
smallest one is smaller than $M_\recomb$, the photon is combined with
the corresponding fermion, \ie the momenta of the photon and the
fermion are added and associated with the momentum of the fermion,
and the photon is discarded.
\item Finally, all events are discarded 
  in which one of the
  final-state fermions is within a cone of 10 degrees around the
  beams.  No other cuts are applied.
\end{enumerate}
We consider the cases of a tight recombination cut $M_\recomb= 5\GeV$,
of a loose recombination cut $M_\recomb= 25\GeV$,
and the case of full photon recombination, $M_\recomb=500\GeV$.
In the following observables, the momenta of the \PW~bosons are always
defined by the sum of the momenta of the two corresponding decay
fermions after the eventual recombination with the photon.

We now discuss various distributions for the $\nu_\mu\mu^+\Pd\Pubar$
final state. These results have been obtained from 50 million events.
In the following figures we always show on the left-hand side the
absolute distributions in lowest order and including the corrections
for the recombination cut $M_\recomb=5\GeV$, and on the right-hand
side the corresponding relative corrections for the three
recombination cuts $M_\recomb=5\GeV$, $M_\recomb=25\GeV$, and
$M_\recomb=500\GeV$.

\begin{figure}
\centerline{%
\setlength{\unitlength}{1cm}
\begin{picture}(7.9,8.2)
\put(0,0){\includegraphics{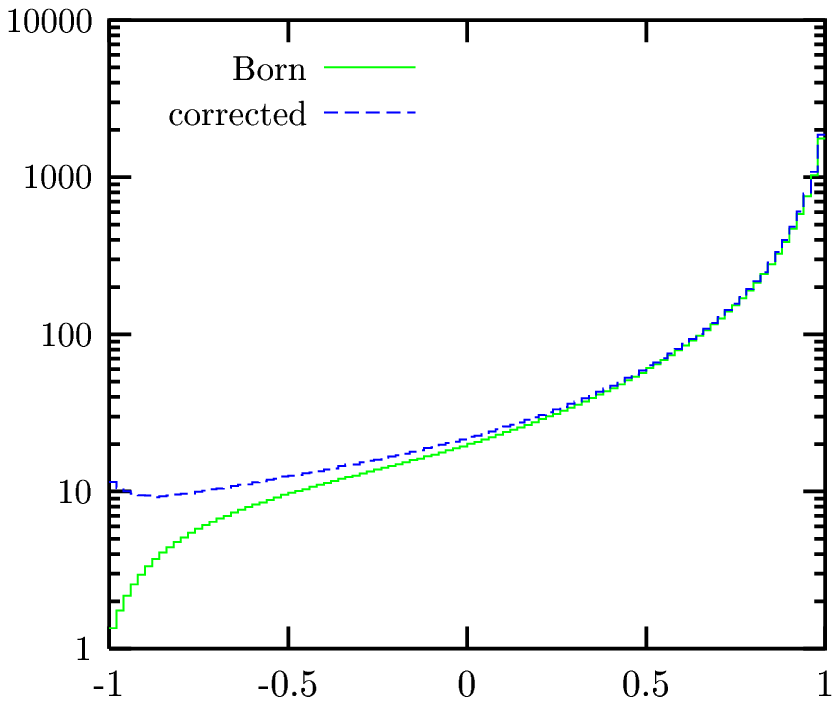}}
\put(0.1,7.3){\makebox(1,1)[l]{$\frac{\rd \si}{\rd \cos\theta_{\PW}}\ 
\left[\fb\right]$}}
\put(4.0,-0.2){\makebox(1,1)[c]{$\cos\theta_{\PW}$}}
\end{picture}%
\begin{picture}(7.9,8.2)
\put(0.5,7.3){\makebox(1,1)[c]{$\de\ [\%]$}} 
\put(4.0,-0.2){\makebox(1,1)[c]{$\cos\theta_{\PW}$}}
\put(0,0){\includegraphics{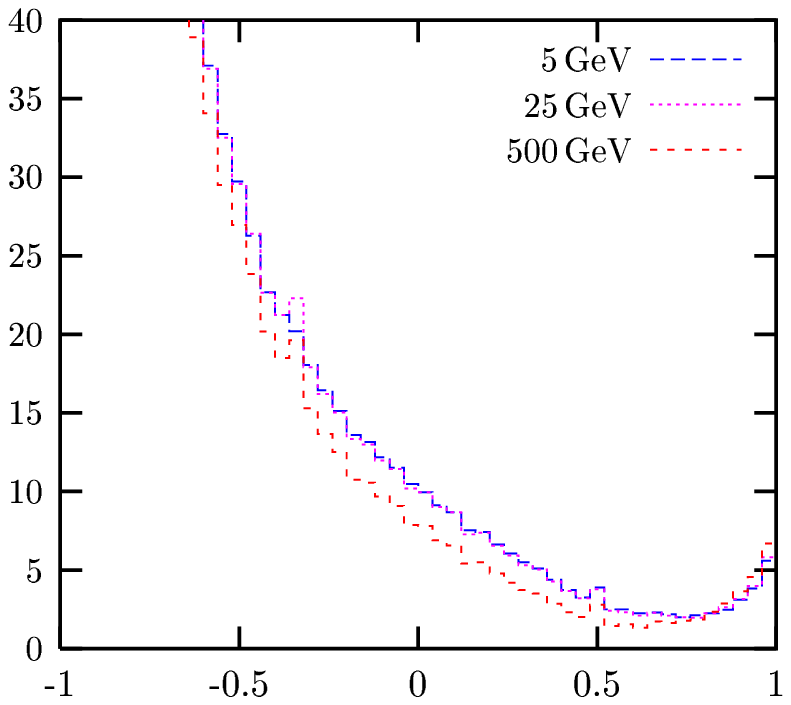}}
\end{picture}
}
\caption{Production-angle distribution for
  $\Pep\Pem\to\nu_\mu\mu^+\Pd\Pubar$ and $\protect\sqrt{s}=500\GeV$}
\label{fi:prod_angle}
\end{figure}

In \reffi{fi:prod_angle} we present the distribution of events in the
angle between the $\PWp$ and the incoming $\Pep$. The corrections lead
to a drastic increase of the cross section in the backward direction.
This effect is due to hard-photon emission from the initial state,
which boosts the CM-system of the W bosons and causes a migration of
events from regions with large cross section in the CM system to
regions with small cross section in the laboratory system.  The
relative
corrections are very large there, because the lowest-order cross
section is very small.  Owing to the small cross section, this region
of phase space is rather unimportant.  Similar but less dramatic boost
effects are also visible in the following distributions. The
production-angle distribution depends on the recombination scheme only
weakly, as expected.  An increase of the recombination cut leads to a
redistribution of events from the backward to the forward direction.
Since there is more phase space for hard photons for higher scattering
energies, the effects of a full recombination are more pronounced at
$\sqrt{s}=500\GeV$ than at LEP2 energies, discussed in
\citere{racoonww1}.  This feature can also be observed in the
following distributions.

\begin{figure}
\centerline{%
\setlength{\unitlength}{1cm}
\begin{picture}(7.9,8.2)
\put(0,0){\includegraphics{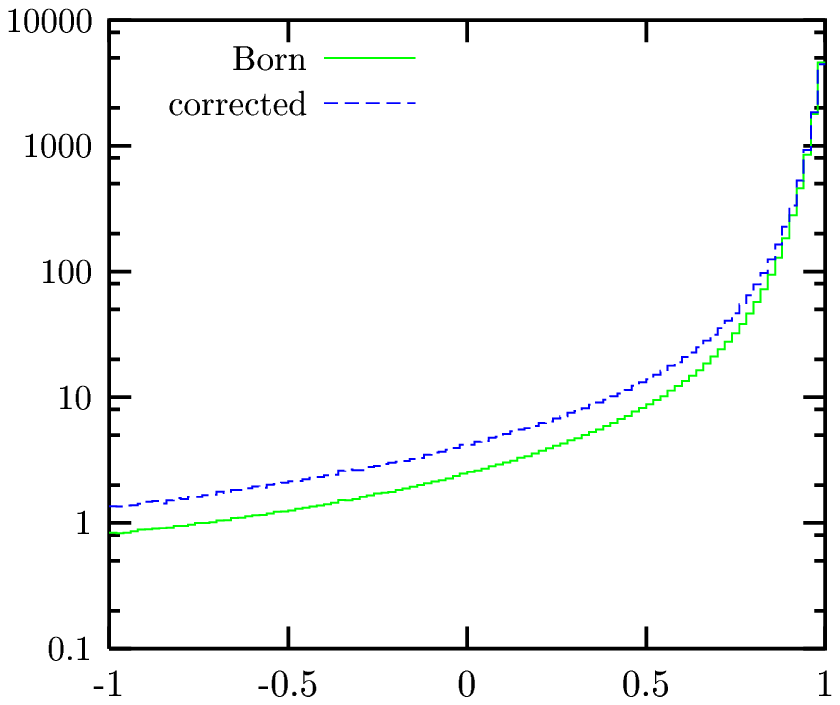}}
\put(0.1,7.3){\makebox(1,1)[l]{$\frac{\rd \si}{\rd \cos\theta_{\PWp\mu^+}}\ 
\left[\fb\right]$}}
\put(4.0,-0.2){\makebox(1,1)[c]{$\cos\theta_{\PWp\mu^+}$}}
\end{picture}%
\begin{picture}(7.9,8.2)
\put(0.5,7.3){\makebox(1,1)[c]{$\de\ [\%]$}} 
\put(4.0,-0.2){\makebox(1,1)[c]{$\cos\theta_{\PWp\mu^+}$}}
\put(0,0){\includegraphics{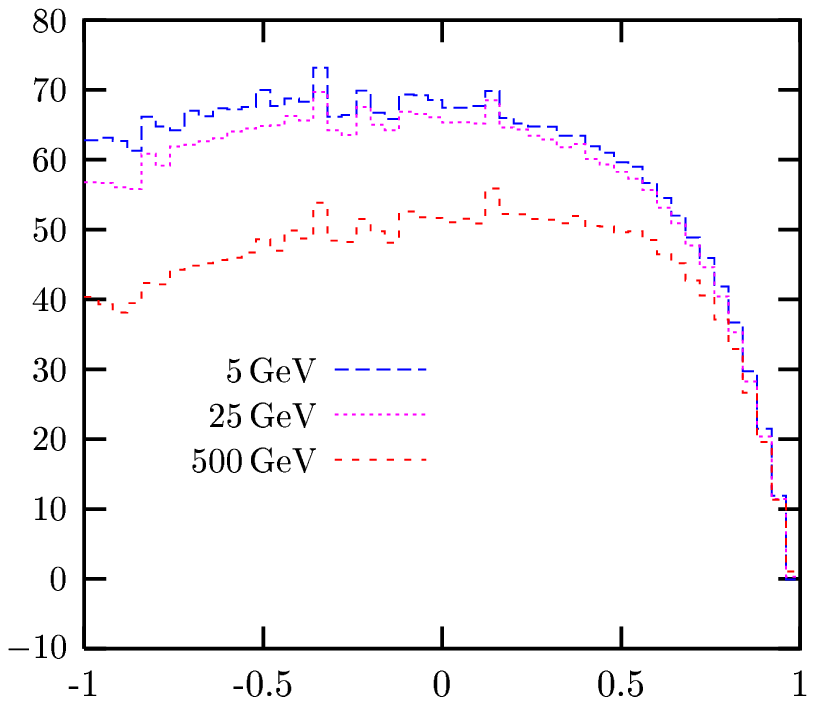}}
\end{picture}
}
\caption{Decay-angle distribution for
  $\Pep\Pem\to\nu_\mu\mu^+\Pd\Pubar$ and $\protect\sqrt{s}=500\GeV$}
\label{fi:decay_angle}
\end{figure}

The distribution of events in the angle between the $\PWp$ and the
outgoing $\mu^+$ is presented in \reffi{fi:decay_angle}.  In this case
we find a noticeable dependence on the recombination mass $M_\recomb$
for large decay angles, where the cross section is, however, very
small.  This originates from the fact that the recombination of a
fermion with a photon parallel to this fermion decreases the angle
between the fermion and the W~boson from which the fermion results.

\begin{figure}
\centerline{%
\setlength{\unitlength}{1cm}
\begin{picture}(7.9,8.2)
\put(0,0){\includegraphics{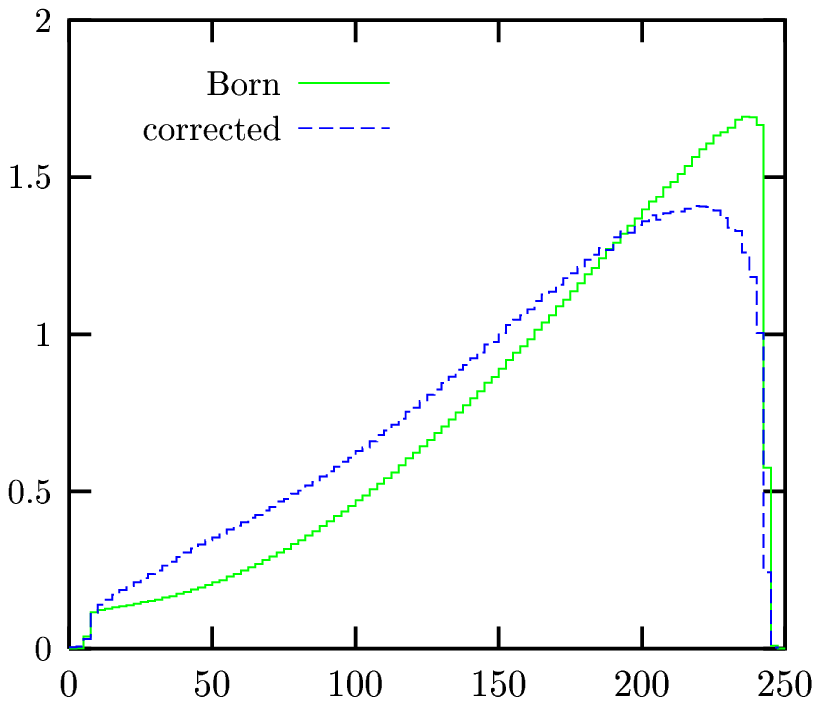}}
\put(0.1,7.3){\makebox(1,1)[l]{$\frac{\rd \si}{\rd E_\mu}\ 
\left[\frac{\fb}\GeV\right]$}}
\put(4.0,-0.2){\makebox(1,1)[c]{$E_\mu\ [\mathrm{GeV}]$}}
\end{picture}%
\begin{picture}(7.9,8.2)
\put(0.5,7.3){\makebox(1,1)[c]{$\de\ [\%]$}} 
\put(4.0,-0.2){\makebox(1,1)[c]{$E_\mu \ [\mathrm{GeV}]$}}
\put(0,0){\includegraphics{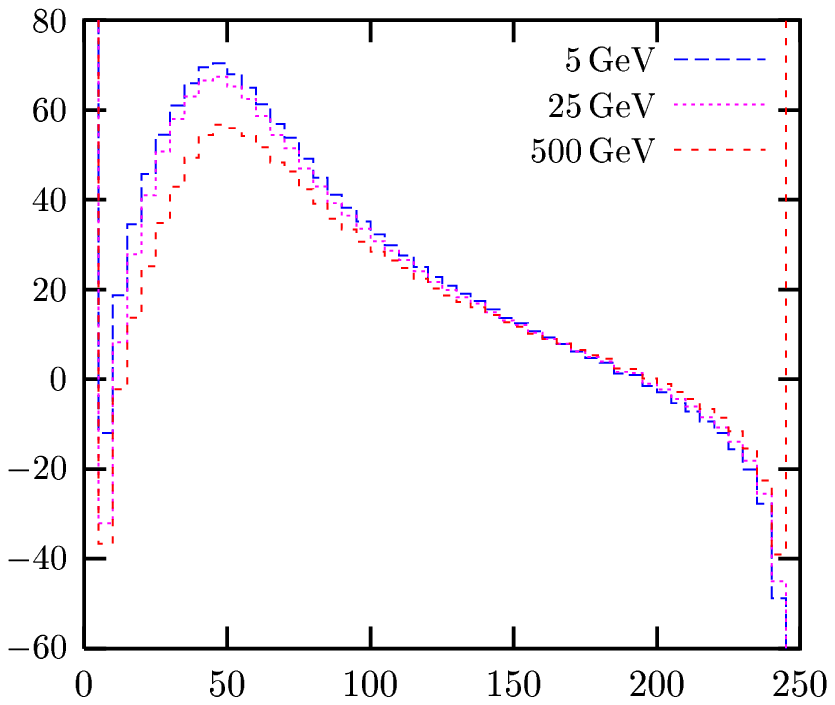}}
\end{picture}
}
\caption{Muon-energy distribution for
  $\Pep\Pem\to\nu_\mu\mu^+\Pd\Pubar$ and $\protect\sqrt{s}=500\GeV$}
\label{fi:mu_energy}
\end{figure}

The distribution of events in the energy $E_\mu$ of the outgoing
$\mu^+$ is depicted in \reffi{fi:mu_energy}.  In the on-shell
approximation it would be restricted between
$6.6\GeV<E_\mu<243.4\GeV$.  Again the radiative corrections tend to
smooth the distribution.  The recombination of a photon with the muon
increases the muon energy. Consequently, an increase of the
recombination mass $M_\recomb$ shifts the distribution to larger muon
energies, 
as can be seen in the relative corrections.

\begin{figure}
\centerline{%
\setlength{\unitlength}{1cm}
\begin{picture}(7.9,8.2)
\put(0,0){\includegraphics{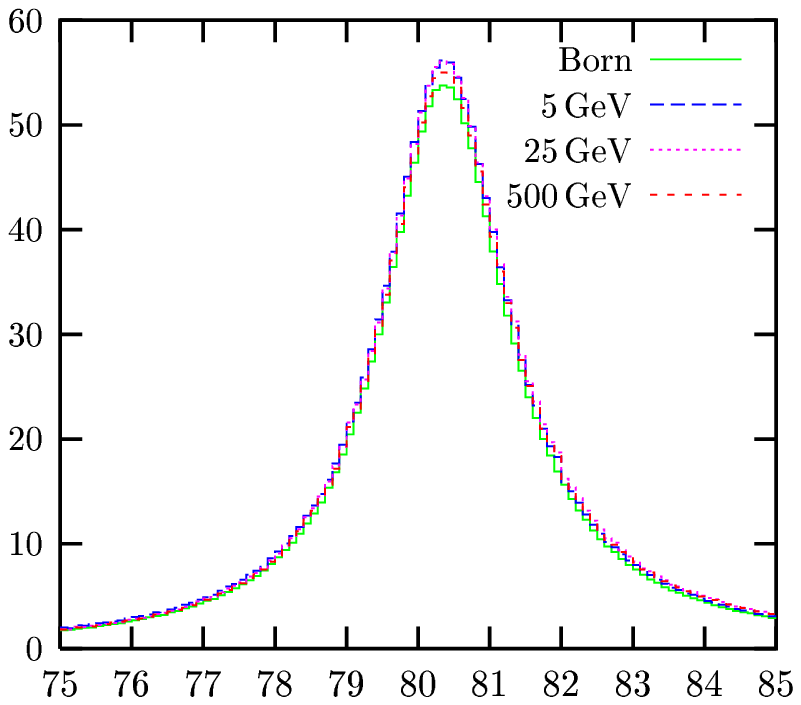}}
\put(0.1,7.3){\makebox(1,1)[l]{$\frac{\rd \si}{\rd M_{\mu\nu_\mu}}\ 
\left[\frac{\fb}\GeV\right]$}}
\put(4.0,-0.2){\makebox(1,1)[c]{$M_{\mu\nu_\mu}\ [\mathrm{GeV}]$}}
\end{picture}%
\begin{picture}(7.9,8.2)
\put(0.5,7.3){\makebox(1,1)[c]{$\de\ [\%]$}} 
\put(4.0,-0.2){\makebox(1,1)[c]{$M_{\mu\nu_\mu}\ [\mathrm{GeV}]$}}
\put(0,0){\includegraphics{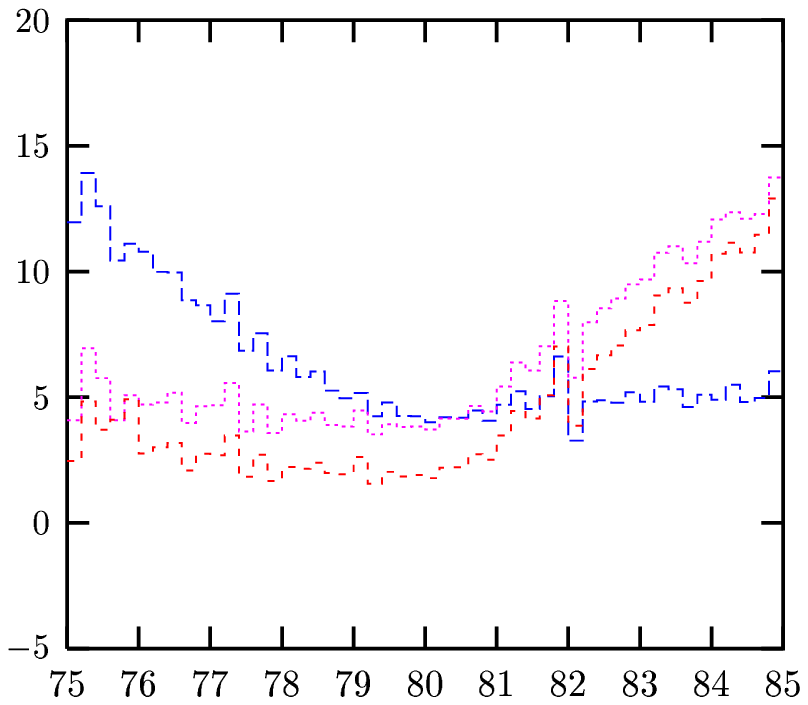}}
\end{picture}
}
\caption{Invariant-mass distribution of the lepton pair for
  $\Pep\Pem\to\nu_\mu\mu^+\Pd\Pubar$ and $\protect\sqrt{s}=500\GeV$}
\label{fi:munu_invmass}
\end{figure}

\begin{figure}
\centerline{%
\setlength{\unitlength}{1cm}
\begin{picture}(7.9,8.2)
\put(0,0){\includegraphics{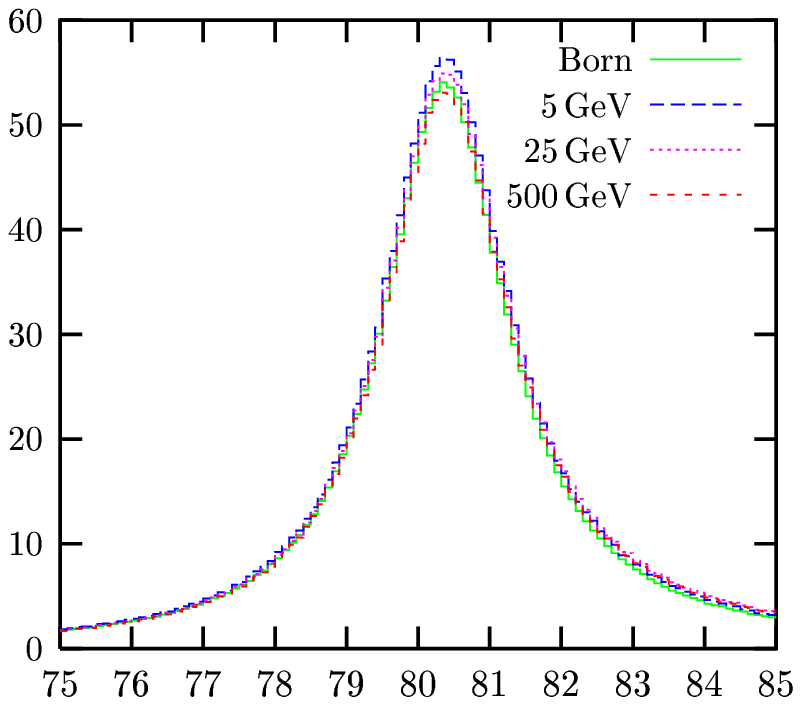}}
\put(0.1,7.3){\makebox(1,1)[l]{$\frac{\rd \si}{\rd M_{\Pd\Pu}}\ 
\left[\frac{\fb}\GeV\right]$}}
\put(4.0,-0.2){\makebox(1,1)[c]{$M_{\Pd\Pu}\ [\mathrm{GeV}]$}}
\end{picture}%
\begin{picture}(7.9,8.2)
\put(0.5,7.3){\makebox(1,1)[c]{$\de\ [\%]$}} 
\put(4.0,-0.2){\makebox(1,1)[c]{$M_{\Pd\Pu}\ [\mathrm{GeV}]$}}
\put(0,0){\includegraphics{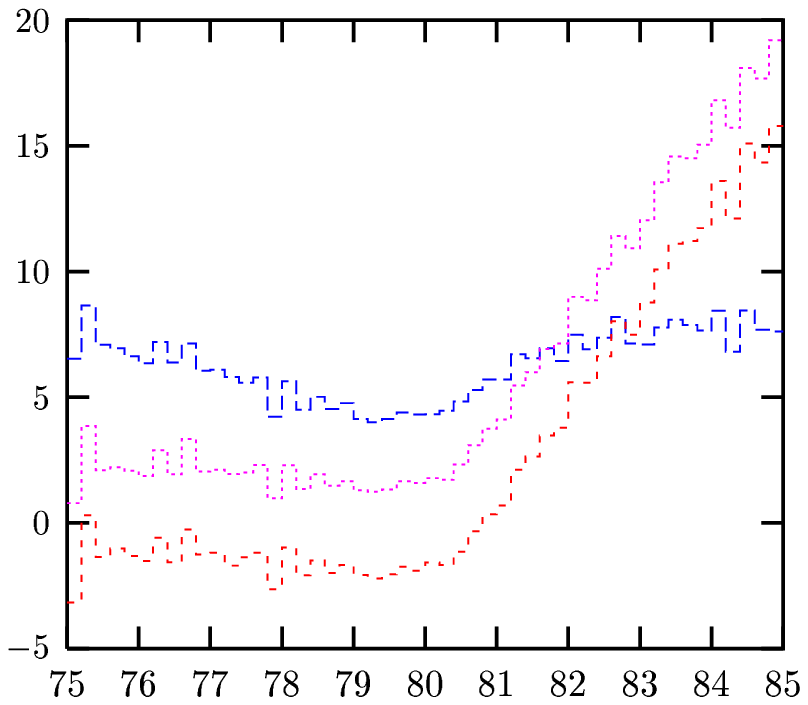}}
\end{picture}
}
\caption{Invariant-mass distribution of the quark pair for
  $\Pep\Pem\to\nu_\mu\mu^+\Pd\Pubar$ and $\protect\sqrt{s}=500\GeV$}
\label{fi:ud_invmass}
\end{figure}

Finally, in \reffis{fi:munu_invmass} and \ref{fi:ud_invmass} we show 
the distributions of events in the invariant masses of the final-state
lepton pair, $M_{\mu\nu_\mu}$, and of the final-state quark pair,
$M_{\Pd\Pu}$. The dependence on the recombination cut is
sizeable everywhere.
The results for the invariant-mass distributions can be understood as
follows. For small recombination cuts, in most of the events the
W~bosons are defined from the decay fermions only. If a photon is
emitted from the decay fermions
and not recombined, the invariant mass of the fermions is smaller than
the one of the decaying W~boson. This leads to an enhancement of the
distribution for invariant masses below the \PW~resonance. This effect
becomes smaller with increasing recombination mass.  The enhancement
is proportional to the squared charges of the final-state fermion, \ie
it is largest for the leptonic invariant mass. On the other hand, if
the recombination mass gets large,
the probability increases that the recombined fermion momenta receive
contributions from photons that are radiated during the W-production
subprocess or from the decay fermions of the other W~boson.
This leads to positive corrections above the considered \PW~resonance.
The effect is larger for the hadronic invariant mass since in this
case, two decay fermions (the two quarks) can be combined with the
photon. The effect of the squared charges of the final-state fermions
is marginal in this case because the contribution of initial-state
fermions dominates.

In contrast to LEP2 energies \cite{racoonww1} there is a significant
difference between the loose recombination cut $M_\recomb=25\GeV$ and
the full recombination $M_\recomb=500\GeV$. As already explained in
\citere{racoonww1},
the recombination of a photon that forms an invariant mass
of at least $25\GeV$ with charged fermions
shifts the events by at least $3.8 \GeV$ for the leptonically-decaying
\PW~boson and $7.4\GeV$ for the hadronically-decaying \PW~boson. Such
a shift leads to an increase of the corrections only outside the range
shown in \reffis{fi:munu_invmass} and \ref{fi:ud_invmass}. 
However, in contrast to the LEP2 energy, 
the phase space for hard photons is very large at $\sqrt{s}=500\GeV$,
leading to a global significant decrease of the W~line shape near the
resonance if the recombination cut $M_\recomb$ increases from $25\GeV$ 
to $500\GeV$.

The distortion of the W~invariant-mass distributions is of particular
interest for the reconstruction of the W-boson mass $\MW$ from the decay
products. In order to illustrate the impact of the corrections on the
determination of $\MW$, we fit the simple Breit--Wigner distribution 
\beq
\biggl(\frac{\rd\sigma}{\rd M^2}\biggr) =
\frac{\mbox{const.}}{(M^2-\MWfit^2)^2+\MWfit^2 \GWfit^2}
\eeq
to the W~line shapes shown in \reffis{fi:munu_invmass} and
\ref{fi:ud_invmass}, treating $\MWfit$ and $\GWfit$ (as well as the
normalization constant) as free fit parameters.  We determine the
fitted \PW-boson masses from the predictions resulting from the
lowest-order CC03 diagrams, $\MWfit^{\born,\mathrm{CC03}}$, from the
complete lowest-order diagrams, $\MWfit^{\born}$, and from the fully
corrected predictions, $\MWfit^{\corr}$. In addition we give the mass
shift resulting from the corrections $\Delta\MWfit^{\corr} =
\MWfit^{\corr} - \MWfit^{\born}$.  The results of these fits, which
are contained in \refta{tablemw}, show that the fitted W-boson mass
changes at the order of some $10\MeV$ if the corrections are included.
The mass shift crucially 
depends on the recombination procedure.
From the discussion of the line-shape distortion
above it is clear that this mass shift is more positive if more
photons are recombined.
Moreover, the differences in $\MWfit^{\corr}$ between the recombination
cuts $M_\recomb=25\GeV$ and $500\GeV$ are only marginal, since the
corresponding line-shape distortions are practically the same.
The results also illustrate that the fit
results vary at the order of some $10\MeV$ for different fit ranges 
in $M$.
\begin{table}
$$
\begin{array}{c@{\qquad}c@{\qquad}c@{\qquad}c@{\qquad}c@{\qquad}c}
\hline
& M_\recomb &
\MWfit^{\born,\mathrm{CC03}\strut} & \MWfit^{\born} &  
\MWfit^{\corr} & \Delta\MWfit^{\corr} \nl
& [\mathrm{GeV}] &
[\mathrm{GeV}] & [\mathrm{GeV}] &  
[\mathrm{GeV}]  & [\mathrm{MeV}] \nl
\hline
\PWp\to\nu_\mu \mu^+ &  5 & 80.377 & 80.376 & 80.377 &  +1 \nl
                     &    & 80.388 & 80.386 & 80.386 &  +0 \nl
                     & 25 & 80.377 & 80.376 & 80.389 & +13 \nl
                     &    & 80.388 & 80.386 & 80.405 & +19 \nl
                     &500 & 80.377 & 80.376 & 80.389 & +13 \nl
                     &    & 80.388 & 80.386 & 80.405 & +19 \nl
\hline
\PWm\to\Pd\Pubar     &  5 & 80.378 & 80.377 & 80.389 & +11 \nl
                     &    & 80.388 & 80.388 & 80.402 & +14 \nl
                     & 25 & 80.378 & 80.377 & 80.401 & +23 \nl
                     &    & 80.388 & 80.388 & 80.422 & +35 \nl
                     &500 & 80.378 & 80.377 & 80.401 & +24 \nl
                     &    & 80.388 & 80.388 & 80.424 & +36 \nl
\hline
\end{array}
$$
\caption{Results for $\MWfit$ for a Breit--Wigner fit to the 
invariant-mass distributions shown in \reffis{fi:munu_invmass} and 
\ref{fi:ud_invmass}, using the fit ranges 
$78.3\GeV < M < 82.3\GeV$ (upper values) and
$76.3\GeV < M < 84.3\GeV$ (lower values) }
\label{tablemw}
\end{table}
\bigskip

In summary, we have applied the event generator {\sc RacoonWW} for
$\eeWWffff(+\gamma)$, which includes the complete $\Oa$ electroweak
corrections in double-pole approximation, for the linear-collider
centre-of-mass energy $500\GeV$.
With this generator we have calculated the total cross sections and
various distributions of experimental relevance.  For angular and
energy distributions we find corrections up to several $10\%$, i.e.\ 
they are much larger than at LEP2.  Since the bulk of these
corrections is due to a redistribution of events by kinematical
effects induced by hard-photon radiation, corrections of even higher
order are expected to be less drastic.  For invariant-mass
distributions of the W~bosons, the size of the corrections is similar
as at LEP2.  The detailed numerical discussion demonstrates that the
issue of photon recombination is of great importance in view of the
aimed accuracy of $15\MeV$ in the kinematical reconstruction of $\MW$
at TESLA \cite{ecfa}.

\end{document}